\documentclass[%
 reprint,
superscriptaddress,
 amsmath,amssymb,
 aps,
floatfix,
]{revtex4-2}

\usepackage{graphicx}% Include figure files
\usepackage{dcolumn}% Align table columns on decimal point
\usepackage{bm}% bold math
\usepackage{times}
\usepackage[usenames,dvipsnames]{xcolor}
\usepackage[colorlinks=true, linkcolor=Blue, citecolor=Blue, urlcolor=Blue]{hyperref} 
\usepackage{scalerel}
\usepackage{stackengine}

\newcommand\reallywidetilde[1]{\ThisStyle{%
  \setbox0=\hbox{$\SavedStyle#1$}%
  \stackengine{-.1\LMpt}{$\SavedStyle#1$}{%
    \stretchto{\scaleto{\SavedStyle\mkern.2mu\sim}{.5467\wd0}}{.7\ht0}%
  }{O}{c}{F}{T}{S}%
}}

\usepackage{color}

\begin{document}

\preprint{APS/123-QED}

\title{Capacitively Shunted Double-Transmon Coupler Realizing Bias-Free Idling and High-Fidelity CZ Gate}% Force line breaks with \\

% Rui Li, Kentaro Kubo, Yinghao Ho, ZhiGuang Yan, Yasunobu Nakamura, and Hayato Goto

\author{Rui Li}
\email{rui.li.gj@riken.jp}
\affiliation{RIKEN Center for Quantum Computing (RQC), Wako, Saitama 351-0198, Japan}
\author{Kentaro Kubo}
\affiliation{Frontier Research Laboratory, Corporate Research \& Development Center,Toshiba Corporation, Saiwai-ku, Kawasaki 212-8582, Japan}
\author{Yinghao Ho}
\affiliation{Frontier Research Laboratory, Corporate Research \& Development Center,Toshiba Corporation, Saiwai-ku, Kawasaki 212-8582, Japan}
\author{Zhiguang Yan}
\affiliation{RIKEN Center for Quantum Computing (RQC), Wako, Saitama 351-0198, Japan}
\author{Shinichi Inoue}
\affiliation{Department of Applied Physics, Graduate School of Engineering, The University of Tokyo, Bunkyo-ku, Tokyo 113-8656, Japan}
\author{Yasunobu Nakamura}
\email{yasunobu@ap.t.u-tokyo.ac.jp}
\affiliation{RIKEN Center for Quantum Computing (RQC), Wako, Saitama 351-0198, Japan}
\affiliation{Department of Applied Physics, Graduate School of Engineering, The University of Tokyo, Bunkyo-ku, Tokyo 113-8656, Japan}
\author{Hayato Goto}
\email{hayato1.goto@toshiba.co.jp}
\affiliation{Frontier Research Laboratory, Corporate Research \& Development Center,Toshiba Corporation, Saiwai-ku, Kawasaki 212-8582, Japan}
\affiliation{RIKEN Center for Quantum Computing (RQC), Wako, Saitama 351-0198, Japan}

% \collaboration{MUSO Collaboration}%\noaffiliation

%  \homepage{http://www.Second.institution.edu/~Charlie.Author}
% \affiliation{
%  Second institution and/or address\\
%  This line break forced% with \\
% }%
% \affiliation{
%  Third institution, the second for Charlie Author
% \affiliation{%
%  Authors' institution and/or address\\
%  This line break forced with \textbackslash\textbackslash
% }%

% \collaboration{CLEO Collaboration}%\noaffiliation

\date{\today}% It is always \today, today,

\begin{abstract}
A high-fidelity CZ gate utilizing a double-transmon coupler~(DTC) has recently been demonstrated as a building block for superconducting quantum processors. Like many other kinds of tunable couplers, however, the DTC requires a finite DC current for flux-biasing the coupler at the idling point to turn off the coupling, necessitating extra care for wiring and heat-load management. To address this issue, we theoretically propose and experimentally realize a novel coupling scheme by introducing a shunt capacitance between the two transmons of the DTC at zero-flux bias, which demonstrates high-fidelity CZ-gate performance comparable to the previous DTC. Through a comprehensive error budget analysis using multiple randomized benchmarking methods, we also identify that the current fidelity is limited by the decoherence through the coupler. Moreover, we experimentally demonstrate the wide operational flux range of the capacitively shunted DTC, which solves the challenging issue of remnant flux existing even with careful magnetic shielding.
\end{abstract}

% Through a comprehensive error budget analysis using multiple randomized benchmarking methods, we identified a nonzero error bound for the incoherent error in the CZ gate. This finding can be quantitatively attributed to the impact of the coupler's decoherence.

\maketitle

Achieving high-fidelity gates is critical for enabling quantum advantage in practical quantum computers. Consequently, developing a highly stable and easily controllable coupling scheme for high-fidelity gates~\cite{PhysRevLett.127.080505,PhysRevX.14.041050,PRXQuantum.4.010314,PhysRevX.13.031035} has emerged as a key challenge, drawing significant attention. Particularly for the widely used superconducting transmon qubits~\cite{koch2007charge, PhysRevLett.111.080502}, a diverse range of coupling schemes has been developed, employing both fixed~\cite{PhysRevLett.94.240502, PhysRevLett.107.080502, PhysRevLett.117.250502, PhysRevLett.130.260601} and tunable~\cite{PhysRevApplied.6.064007, PhysRevApplied.10.054062, PhysRevX.14.041050, PRXQuantum.4.010314, PhysRevLett.127.080505, PhysRevApplied.20.044028} coupling modalities. Compared to the cases with fixed coupling, the mediated qubit--qubit interaction through a  tunable coupler can be switched on and off with a higher contrast using either a microwave drive or baseband pulse control~\cite{ding2024pulsedesignbasebandflux}. This typically enables faster two-qubit gates, resulting in higher gate fidelity~\cite{PhysRevA.93.060302, PhysRevX.11.021058}.

Among various tunable couplers, the recently proposed double-transmon coupler~(DTC)~\cite{PhysRevApplied.18.034038, kubo2023fast, PhysRevApplied.19.064043, PhysRevApplied.22.024057} has experimentally demonstrated its advantage in enabling high fidelities for both single-qubit and CZ gates~\cite{PhysRevX.14.041050}. 
Similarly to other tunable couplers~\cite{PRXQuantum.4.010314, PhysRevLett.127.080505, PhysRevX.11.021058}, however, the DTC requires a large DC flux bias~($\sim$0.3 flux quantum) to suppress residual interactions between qubits, the so-called \textit{ZZ} interaction~\cite{PhysRevA.102.022619}, for idling and to enable high-fidelity gates. 
In large-scale quantum processors, wiring and heat-load management for the individual DC current biasing are nontrivial. Therefore, bias-free couplers are desired for further scale-up.

In this work, we theoretically propose and experimentally realize a DC-bias-free DTC scheme by introducing a large coupling capacitance shunting the two coupler transmons, referred to as the capacitively shunted double-transmon coupler (CSDTC). We first explain why the \textit{ZZ}
interaction can be suppressed by the shunt capacitance even at the zero flux bias. Then, we demonstrate that a negligible \textit{ZZ} interaction can be achieved over a wide flux-bias range around the zero-flux point. 
This property is important for bias-free tunable couplers, because remnant stray flux within the chip package, even with adequate shielding, is non-negligible in actual experiments.
Furthermore, by activating the CZ gate with a net-zero~(NZ) pulse robust against low-frequency flux fluctuations~\cite{PhysRevLett.123.120502, PhysRevApplied.16.054020, PhysRevLett.126.220502}, we employ leakage randomized benchmarking~(LRB)~\cite{PhysRevX.14.041050} and purity randomized benchmarking ~(PRB)~\cite{PhysRevLett.117.260501, wallman2015estimating} to characterize various CZ-gate errors, including the leakage, coherent, and incoherent errors. By varying the gate duration, we show that the CZ-gate fidelities are dominated by the incoherent error through the influence of decoherence effects associated with the coupler modes. Finally, by intentionally biasing the coupler flux within the range with negligible \textit{ZZ} interaction, we demonstrate that the CSDTC enables high fidelities for both single-qubit and CZ gates even in the presence of such a fictitious stray flux. This result highlights its broad operational range, eliminating the need for a DC bias to set the coupler to a specific idle point. The high-fidelity CZ gates also benefit from the use of a biased net-zero~(BNZ) pulse, which is robust to flux noise and pulse distortion, achieving a CZ-gate fidelity of 99.89\%, comparable to the previous DTC study~\cite{PhysRevX.14.041050}.

\begin{figure}
\includegraphics[scale=1.0]{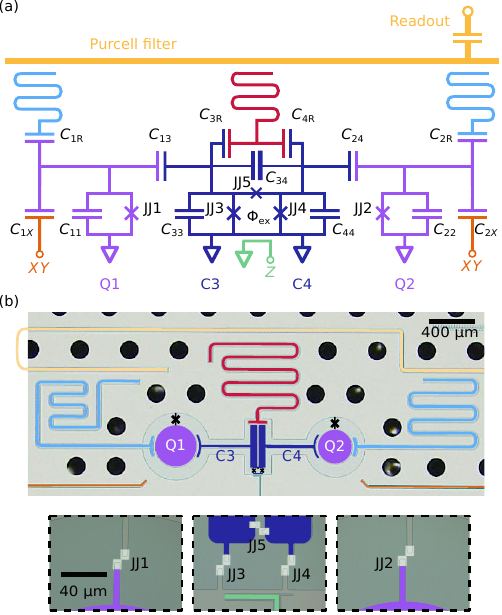}% Here is how to import EPS art
\caption{\label{Fig1:wide} (a)~Circuit diagram of the device, where two transmon qubits, Q1 and Q2, are coupled via the CSDTC. The CSDTC consists of two coupler transmons, C3 and C4, which are coupled through a Josephson junction JJ5 and a capacitance $C_{34}$ in parallel. Dedicated control and readout lines for the qubits and the CSDTC are also illustrated. (b)~False-color picture of the device. The black circles are superconducting through-silicon vias (TSVs) fabricated throughout the device to suppress spurious modes by connecting the ground planes on the top and bottom surface of the chip. The Josephson-junction sections are magnified in the bottom panels.} 
\end{figure}
% This file may be formatted in either the \texttt{preprint} or

The mechanism of turning off the residual \textit{ZZ} interaction without using flux bias in the CSDTC is explained qualitatively as follows.
The circuit diagram of the CSDTC is shown at the center of Fig.~\ref{Fig1:wide}(a) (dark-blue part), 
where two fixed-frequency transmons, C3 and C4, are coupled by a Josephson junction JJ5 with Josephson energy $E_\mathrm{J5}$ and a shunt capacitance $C_{34}$. 
When the flux through the coupler loop, consisting of junctions JJ3, JJ4, and JJ5, is zero, JJ5 can be approximated well as an inductor with the Josephson inductance $L_{\mathrm{J}5}=[\Phi_0/(2\pi)]^2/E_\mathrm{J5}$, 
where $\Phi_0 = h/(2e)$ is the flux quantum and $h$ and $e$ are the Planck constant and elementary charge, respectively. 
Then, the composite impedance $Z_{\mathrm{C}}(\omega )$ of $C_{34}$ and $L_{\mathrm{J}5}$ is given by
\begin{align}
    Z_\mathrm{C}(\omega) = \left(i\omega C_{34} + \frac{1}{i\omega L_{\mathrm{J}5}}\right)^{\! -1}.
\end{align}
Note that this diverges at ${\omega = 1/\sqrt{L_{\mathrm{J}5} C_{34}}}$.
Therefore, we can almost turn off the interaction between qubits Q1 and Q2
by choosing the shunt capacitance as ${C_{34} = 1/(L_{\mathrm{J}5} \omega^2)}$, 
where $\omega$ is set around the qubit frequencies $\omega_1$ and $\omega_2$. 
More accurate analyses based on classical and quantum mechanics show that an appropriate value of $\omega$ is the geometric mean $\sqrt{\omega_1 \omega_2}$~\cite{SOM}.

To demonstrate the advantage of the CSDTC scheme, we devise a circuit shown in Fig.~\ref{Fig1:wide} that utilizes a CSDTC to couple two data transmons. The external flux penetrating through the coupler $\Phi_\mathrm{ex}$ can be controlled either through an adjacent fast flux line or a global flux coil. 
When the coupler is biased at the zero-flux point by applying a global flux compensating the remnant flux, Q1 has a frequency of $3944$~MHz and an anharmonicity of $-175$~MHz, while Q2 has a frequency of $4438$~MHz and an anharmonicity of $-208$~MHz. The two data qubits are fixed-frequency transmons. However, owing to the strong coupling between the qubits and the coupler, the qubit frequencies can shift depending on the flux bias to the coupler loop. 
 The large $C_{34}$ is achieved through the direct mutual capacitance between the coupler transmons, C3 and C4~[Figs.~\ref{Fig1:wide}(b)].

\begin{figure}
\includegraphics[scale=1.0]{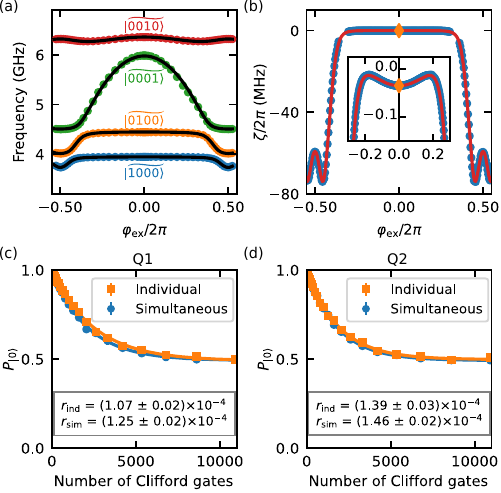}% Here is how to import EPS art
\caption{\label{Fig2:wide} (a)~Energy levels of the lowest four eigenstates as a function of the reduced flux $\varphi_\mathrm{ex}$. (b)~\textit{ZZ} interaction $\zeta$. In (a) and~(b), the data points represent the experimental results, while the solid lines are calculated from the diagonalization of the Hamiltonian in~Eq.~\eqref{basic Hamiltonian}. A zoomed-in view of the \textit{ZZ} interaction near the zero-flux point is shown in the inset of~(b). The point marked by a yellow diamond indicates a \textit{ZZ} interaction of $-35.4$~kHz at the zero-flux point. (c) and (d)~Individual and simultaneous randomized benchmarking of single-qubit gates on Q1 and Q2 at zero-flux point, respectively.
}
\end{figure}

Considering only the two data qubits and the CSDTC, the Hamiltonian can be written as~\cite{PhysRevApplied.18.034038, PhysRevX.14.041050, SOM} 
\begin{equation}\label{basic Hamiltonian}
\begin{aligned}
\hat{H}= &~ \frac{4e^2}{2} \hat{\mathbf{n}}^\mathrm{T}  \mathsf{C}^{-1} \hat{\mathbf{n}} -\sum_{i=1}^4 E_{\mathrm{J}i} \cos \hat{\varphi}_i \\
&- E_{\mathrm{J}5} \cos \left(\hat{\varphi}_4-\hat{\varphi}_3-\varphi_{\mathrm{ex}}\right),
\end{aligned}
\end{equation}
where $\hat{n}_i$ is the Cooper-pair number operator and $\hat{\varphi}_i$ is the node-phase operator, with $i \in \{1, 2, 3, 4\}$ corresponding to the nodes of \{Q1, Q2, C3, C4\}~[Fig.~\ref{Fig1:wide}(a)], and $\hat{\mathbf{n}} = \left\{ \hat{n}_i \right\}$. The capacitance matrix $\mathsf{C}$ is constructed by $\mathsf{C}_{ii} = \sum_{j=1}^4 C_{ij}$ and $\mathsf{C}_{ij} = -C_{ij}$~($i\neq j$). $E_{\mathrm{J}i}$ is the Josephson energy of the Josephson junction JJ$i$, where $i\in \{1, 2, 3, 4, 5\}$ corresponds to the five junctions shown in Fig.~\ref{Fig1:wide}(a). 
The reduced flux $\varphi_{\mathrm{ex}} = 2\pi\Phi_\text{ex}/\Phi_0$ has also been introduced.

%We have dropped the time derivative term of $\varphi_{\mathrm{ex}}$~\cite{PhysRevApplied.18.034038} as it is negligible in our experiment. 
Similarly to the toy model developed in Ref.~\citenum{PhysRevX.14.041050}, the CSDTC contains two low-frequency modes, denoted as P and M. The P mode mimics a fixed-frequency transmon, while the M mode acts as a capacitively shunted flux qubit~(CSFQ)~\cite{yan2016flux}. Following Ref.~\citenum{PhysRevX.14.041050}, we continue using the notation of $|\reallywidetilde{\mathrm{Q1,}\mathrm{Q2,}\mathrm{P,}\mathrm{M}}\rangle$ to represent hybridized states of the two data qubits and the two coupler modes. Based on the energy levels of the four modes measured by varying the external flux, the circuit parameters~\cite{SOM} in the Hamiltonian in~Eq.~\eqref{basic Hamiltonian} are fine-tuned to fit the energy spectra, as depicted in Fig.~\ref{Fig2:wide}(a). 

The \textit{ZZ} interaction $\zeta$, defined as
\begin{equation}\label{ZZ_interaction}
    \hbar\zeta = E_{|\widetilde{1100}\rangle} - E_{|\widetilde{1000}\rangle}  - E_{|\widetilde{0100}\rangle} + E_{|\widetilde{0000}\rangle},
\end{equation}
is then measured~[Fig.~\ref{Fig2:wide}(b)] by using the Joint Amplification of \textit{ZZ} interaction (JAZZ) method~\cite{PhysRevLett.130.260601, PhysRevX.14.041050}, where $E_{|\reallywidetilde{\mathrm{Q1,}\mathrm{Q2,}\mathrm{P,}\mathrm{M}}\rangle}$ denotes the eigenenergies of the four modes. The simulated results match the experimental values well, validating our circuit model. A \textit{ZZ} interaction of $-35.4$~kHz is observed at the zero-flux point, which is confirmed to have a negligible influence on the single-qubit gate fidelities, maintained up to 99.985\%, as verified with simultaneous randomized benchmarking~[Fig.~\ref{Fig2:wide}(c)]. 

Within the flux range of $|\varphi_\mathrm{ex}/2\pi| \leq 0.2$, the \textit{ZZ} interaction $|\zeta/2\pi| \leq 35.4$ kHz~[Fig.~\ref{Fig2:wide}(b)] demonstrates a broad operational range of the CSDTC without requiring any intentional DC bias current, assuming only a moderate remnant flux in the package with adequate shielding. On the other hand, a sufficiently large \textit{ZZ} interaction, on the order of several tens of megahertz, in the flux range of $0.3\leq |\varphi_\mathrm{ex}/2\pi| \leq 0.5$, can be utilized to achieve a fast and high-fidelity CZ gate.

\begin{figure}
\includegraphics[scale=1.0]{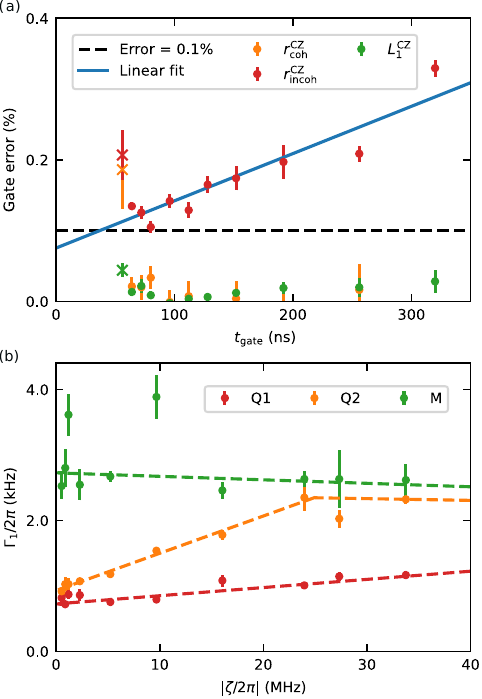}% Here is how to import EPS art
\caption{\label{Fig3:wide} (a)~Error budgets of the CZ gate as a function of its duration $t_\mathrm{gate}$. The solid blue line represents a linear fit to the incoherent errors, excluding the first point, which is marked with a cross to highlight it as an obvious outlier. (b)~Relaxation rates of Q1, Q2, and the M mode as a function of the \textit{ZZ} interaction $\zeta$ between Q1 and Q2. The data points represent experimental results, while the dashed lines serve as guides to the eye.
}
\end{figure}

We first investigate the CZ-gate errors when biasing the CSDTC at the zero-flux point. In addition to carefully mitigating the \textit{Z}-pulse distortion using the cryoscope technique~\cite{rol2020time, PhysRevX.14.041050}, we choose the NZ pulse to activate the CZ gate to further reduce long-time pulse distortion and take advantage of its intrinsic echo effect to extend the coherence time. The NZ pulse is initialized by concatenating two Slepian pulse shapes~\cite{press2007numerical} with opposite poles and then optimized using a model-free reinforcement-learning algorithm~\cite{PhysRevX.14.041050}. By employing the LRB and PRB techniques, we systematically decompose the CZ-gate error into three primary components: incoherent error, coherent error, and leakage error~\cite{SOM}.

\begin{figure*}
\includegraphics[scale=1.0]{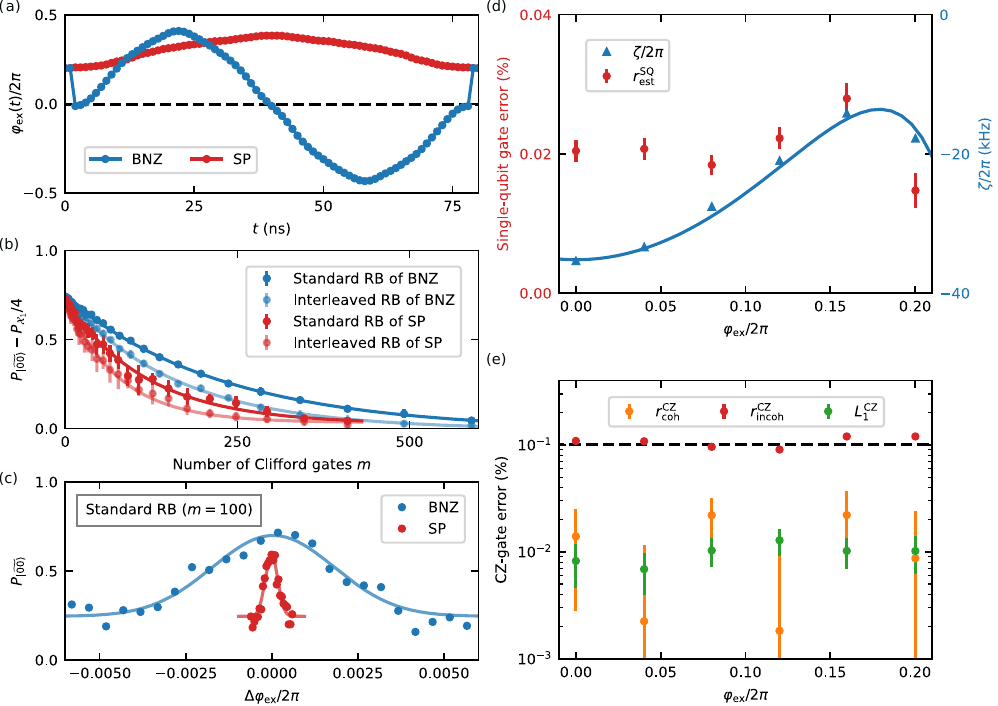}% Here is how to import EPS art
\caption{\label{Fig4:wide} (a)~Typical biased net-zero (BNZ) pulse and single-pole (SP) pulse optimized for achieving a CZ gate in the presence of a finite flux-bias offset~$\varphi_\mathrm{ex}/2\pi = 0.2$. (b)~Interleaved randomized benchmarking results for the CZ gate using optimized BNZ and SP pulses shown in (a), where $\varphi_\mathrm{ex}/2\pi = 0.2$ is applied through a global coil, featuring the experimental data points and the fitting curves. The gate and leakage errors are $r^{\mathrm{CZ}} = 0.14\%$ and $L_1^{{\mathrm{CZ}}} = 0.01\%$ for the BNZ pulse, while $r^{\mathrm{CZ}} = 0.34\%$ and $L_1^{{\mathrm{CZ}}} = 0.01\%$ for the SP pulse. (c)~Population of $|\widetilde{00}\rangle$, $P_{|\widetilde{00}\rangle}$, measured using a standard RB sequence of length 100, as a function of the external flux-bias shift $\Delta \varphi_\mathrm{ex}$ from $\varphi_\mathrm{ex}/2\pi = 0.2$, where the BNZ and SP pulses are calibrated. The experimental data points are fitted to Gaussian curves with the offset determined by the randomization and leakage. (d)~Single-qubit gate errors $r_\mathrm{est}^\mathrm{1Q}$ and measured \textit{ZZ} interaction $\zeta$ against different DC-flux bias $\varphi_\mathrm{ex}$ provided by the global coil. The single-qubit gate errors are estimated from the CZ-gate IRB experiments in (e). The Blue solid curve represents the simulated \textit{ZZ} interactions. (e)~Error budgets of the CZ gate against different DC-flux bias by using an optimized BNZ pulse shape. The coherent and incoherent CZ gate errors, $r_\mathrm{coh}^\mathrm{CZ}$ and $r_\mathrm{incoh}^\mathrm{CZ}$, are separated with the PRB. The  black dashed line indicate a gate error of 0.1\%.
}
\end{figure*}

By varying the gate duration $t_\mathrm{gate}$, we find that the CZ-gate fidelities are primarily limited by incoherent errors~[Fig.~\ref{Fig3:wide}(a)]. Reducing the gate duration is typically an obvious approach to minimize incoherent errors. However, two main factors hinder this strategy. First, all the above three types of errors increase sharply when the gate duration is reduced to 56 ns in our case. The sudden rise in leakage error can be attributed to the non-adiabatic transitions during the CZ pulse, which may result from a limited bandgap between the coupler and qubit modes or suboptimal pulse optimization. The increase of coherent errors could be caused by sequence-dependent phase errors arising from uncalibrated pulse distortions. Additionally, these sequence-dependent phase errors, combined with the randomness of the benchmarking sequence, may introduce significant decoherence during the benchmarking process, thereby increasing the incoherent error. Tuning the frequency of the M mode closer to the qubits is essential to achieve stronger \textit{ZZ} interactions, enabling shorter gate durations. 
As shown in Fig~\ref{Fig3:wide}(b), however, it results in a degradation of qubit coherence~\cite{PRXQuantum.4.010314, PhysRevX.14.041050} and an increase in incoherent errors.

Indeed, by linearly fitting the incoherent error as a function of the gate duration, we observe a non-zero error offset~($r_0 = 0.07\%$) at zero gate duration~[Fig.~\ref{Fig3:wide}(a)]. To quantitatively explain this phenomenon, we model the incoherent error as $r_\mathrm{incoh}^{\mathrm{CZ}} = \Gamma_\mathrm{D}t_\mathrm{gate}$, where $\Gamma_\mathrm{D}$ represents the depolarizing rate during randomized benchmarking. We further observe that the relaxation rates for Q1 and Q2 are correlated with the \textit{ZZ} interaction [Fig.~\ref{Fig3:wide}(b)], which can be attributed to the coupling between the qubits and the M mode. As the M mode approaches the qubits, leading to a stronger \textit{ZZ} interaction, its faster relaxation also leads to an enhanced relaxation rate of the qubits. Within the \textit{ZZ} interaction range of $1\,\mathrm{MHz} < |\zeta/2\pi| < 25\,\mathrm{MHz}$, which approximately corresponds to the gate-duration range investigated, the relaxation rate of each qubit exhibits a linear dependence on the \textit{ZZ} interaction [Fig.~\ref{Fig3:wide}(b)]. This leads to a reasonable assumption of the form $\Gamma_D = a\zeta + k$, which yields $r_\mathrm{incoh}^{\mathrm{CZ}} = (a\zeta + k)t_\mathrm{gate}= r_0 + kt_\mathrm{gate}$ with a constant coefficient $a$ and a $\zeta$-independent incoherent error rate $k$. Here, $r_0 = a\zeta t_\mathrm{gate}$ can be approximated as a nonzero constant, since $\zeta t_\mathrm{gate} \sim \pi$ is required to achieve a $\pi$-phase shift necessary for the CZ gate. We note that a precise estimation of $r_0$ is highly dependent on the state trajectory during the adiabatic CZ gate, as well as the contribution of qubit decoherence to depolarization during the benchmarking process~\cite{PhysRevApplied.16.054020}. However, our result offers a self-consistent, quantitative explanation of how the coherence properties of the coupler influence the fidelity of the CZ gate.

So far, we have discussed the error budget of the CZ gate with the flux bias at the zero-flux point, where the residual \textit{ZZ} interaction is negligible and a NZ pulse can be applied. However, the random distribution of remnant flux within the chip package poses challenges for maintaining the perfect zero-flux bias for large quantum processors with many qubits and couplers without using individual DC bias currents. Here, we demonstrate that the CSDTC can sustain high-fidelity operations across a wide range of flux-bias values, effectively addressing this challenge. In comparison to the conventional single-pole (SP) pulse, we first introduce a biased net-zero~(BNZ) pulse~[Fig.~\ref{Fig4:wide}(a)], which demonstrates improved CZ-gate fidelity at $\varphi_\mathrm{ex}/2\pi = 0.2$~[Fig.~\ref{Fig4:wide}(b)] as an example. This flux bias is applied via a global coil as a constant offset mimicking a remnant flux. In these measurements, we focus exclusively on the states of Q1 and Q2, omitting the coupler-state readout because of the significant crosstalk observed between Q2 and the coupler's readout resonator under a certain current-bias condition. Combined with PRB, our analysis reveals that for the BNZ pulse, the CZ-gate fidelity is primarily limited by the incoherent error. In contrast, for the SP pulse, the CZ-gate fidelity is constrained by the incoherent error and a comparable level of coherent error~\cite{SOM}. In both cases, leakage errors are an order of magnitude lower than incoherent errors~\cite{SOM}. By varying the global flux within a narrow range without recalibrating the CZ-gate pulses, we demonstrate that the standard RB sequence population is almost an order of magnitude more sensitive to the flux-bias shift for the SP pulse compared to the BNZ pulse~[Fig.~\ref{Fig4:wide}(c)]. These results highlight the intrinsic echo effect of the BNZ pulse, which makes it more robust to low-frequency flux noise and pulse distortion~\cite{PhysRevApplied.16.054020}.

% Combined with the PRB, we found that the CZ-gate fidelity for the BNZ pulse is primarily limited by incoherent errors, whereas a large incoherent error emerges for the SP pulse~\cite{SOM}

By varying the global flux within the range $0 \leq \varphi_\mathrm{ex}/2\pi \leq 0.2$, we demonstrate that the CSDTC is effective in achieving high fidelities for both single-qubit~[Fig.~\ref{Fig4:wide}(d)] and BNZ CZ gates~[Fig.~\ref{Fig4:wide}(e)]. We initially characterize the CZ-gate error using IRB and then estimate the single-qubit gate fidelity based on the Clifford-gates decomposition~\cite{barends2014superconducting} in the protocol of randomized benchmarking. The averaged single-qubit gate fidelity obtained in this way accounts for the potential influence of the CZ gate, which is more relevant to practical quantum algorithms. In the CZ-gate error budget, incoherent errors are found to be an order of magnitude larger than leakage and coherent errors across the entire investigated flux range. The CZ-gate fidelity 
given by ${F=1-r^{\mathrm{CZ}}-L_1^{\mathrm{CZ}}/4}$~\cite{PhysRevX.14.041050} exceeds 99.85\% at each flux bias point, with a maximum fidelity of $99.89\pm0.01$\%, primarily attributed to fluctuations in qubit coherence.

In conclusion, we have theoretically proposed and experimentally realized a capacitively-shunted double-transmon-coupler scheme aiming at achieving a high-fidelity CZ gate without DC flux bias for idling. At the zero-flux point, we have evaluated the CZ-gate fidelity using a net-zero pulse and shown that the incoherent error is dominant in the CZ-gate error budget. 
We have also shown quantitatively that the incoherent error is due to the contribution of the coupler's decoherence, revealing a potential direction for developing quantum operations with higher fidelity. The wide flux bias range with negligible \textit{ZZ} interaction in the CSDTC eliminates the need for a DC bias even in the presence of remnant stray flux, making it advantageous for reducing wire congestion and potential crosstalk. With an intentionally added bias offset within the range $|\varphi_\mathrm{ex}/2\pi| \le 0.2$, we demonstrated single-qubit gate fidelities exceeding 99.97\% and CZ-gate fidelities exceeding 99.85\%, highlighting the advantages of using a biased net-zero pulse. Our results demonstrate that the CSDTC is a promising candidate for a building block for future large-scale quantum processors.

\begin{acknowledgments}
We acknowledge K. Kusuyama and Y. Sakoda for the Ta-film deposition; A. Badrutdinov for the TSV\nobreakdash-fabrication assistance. This research was partly funded by the Ministry of Education, Culture, Sports, Science and Technology~(MEXT) Quantum Leap Flagship Program~(Q-LEAP) (Grant No.~JPMXS0118068682).

\end{acknowledgments}

\bibliography{reference}% Produces the bibliography via BibTeX.

\end{document}

	
	%\baselineskip24pt
\title{Supplemental Material for \\ Capacitively Shunted Double-Transmon Coupler Realizing Bias-Free Idling and High-Fidelity CZ Gate}

\author{Rui Li}
\email{rui.li.gj@riken.jp}
\affiliation{RIKEN Center for Quantum Computing (RQC), Wako, Saitama 351-0198, Japan}
\author{Kentaro Kubo}
\affiliation{Frontier Research Laboratory, Corporate Research \& Development Center,Toshiba Corporation, Saiwai-ku, Kawasaki 212-8582, Japan}
\author{Yinghao Ho}
\affiliation{Frontier Research Laboratory, Corporate Research \& Development Center,Toshiba Corporation, Saiwai-ku, Kawasaki 212-8582, Japan}
\author{Zhiguang Yan}
\affiliation{RIKEN Center for Quantum Computing (RQC), Wako, Saitama 351-0198, Japan}
\author{Shinichi Inoue}
\affiliation{Department of Applied Physics, Graduate School of Engineering, The University of Tokyo, Bunkyo-ku, Tokyo 113-8656, Japan}
\author{Yasunobu Nakamura}
\email{yasunobu@ap.t.u-tokyo.ac.jp}
\affiliation{RIKEN Center for Quantum Computing (RQC), Wako, Saitama 351-0198, Japan}
\affiliation{Department of Applied Physics, Graduate School of Engineering, The University of Tokyo, Bunkyo-ku, Tokyo 113-8656, Japan}
\author{Hayato Goto}
\email{hayato1.goto@toshiba.co.jp}
\affiliation{Frontier Research Laboratory, Corporate Research \& Development Center,Toshiba Corporation, Saiwai-ku, Kawasaki 212-8582, Japan}
\affiliation{RIKEN Center for Quantum Computing (RQC), Wako, Saitama 351-0198, Japan}

\maketitle

\section{Device parameters}\label{SUP0}

\begin{figure}[h]
\includegraphics[scale=1.0]{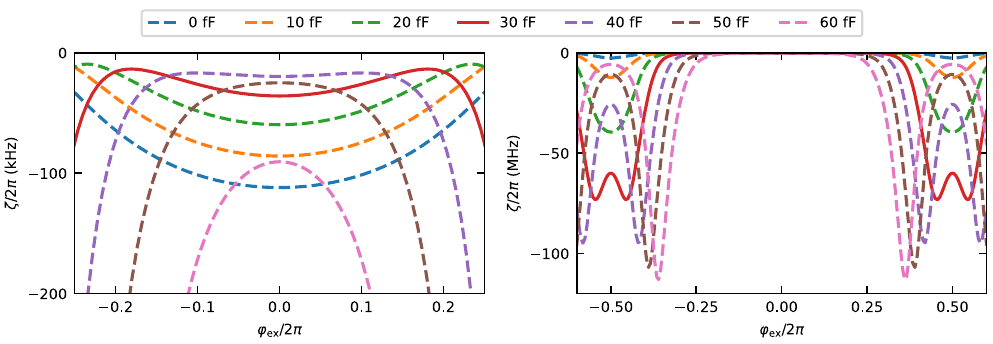}% Here is how to import EPS art
\caption{\label{FigS1:wide} \textit{ZZ} interactions simulated \Imp{with} different $C_{34}$ \Imp{and the} other parameters listed in Table~S\ref{tab:circuit_params}. The red solid line corresponds to the specific condition used in our \Imp{experimental} demonstration.
}
\end{figure}

The circuit parameters, detailed in Table~S\ref{tab:circuit_params}, are utilized to simulate the energy levels and \textit{ZZ} interactions depicted in Fig.~2 of the main text. We also utilize these circuit parameters to simulate \textit{ZZ} interactions by adjusting $C_{34}$~(Fig.~S\ref{FigS1:wide}). In our configuration, with $C_{34} = 30$ fF, we achieve not only a wide range of suppressed \textit{ZZ} interactions around zero flux bias but also maintain a sufficiently large interaction strength at large flux bias to enable fast CZ-gate operations.

\begin{table}[h]
\caption{Circuit parameters for the simulation.}
    \centering
    \begin{tabular}{cccccc}
        \hline\hline \text { Capacitance }& & \text { $C_{11}$ } & \text { $C_{22}$ } & \text { $C_{33}$ } & \text { $C_{44}$}  \\
        \hline \text { Value (fF) } & & 108 & 80 & 90 & 90\\
        \hline\hline
    \end{tabular}

     \begin{tabular}{ccccccc}
        \hline\hline \text { Critical current }& & \text { $I_{\mathrm{c}1}$ } & \text { $I_{\mathrm{c}2}$ } & \text { $I_{\mathrm{c}3}$ } & \text { $I_{\mathrm{c}4}$} & \text { $I_{\mathrm{c}5}$}  \\
        \hline \text { Value (nA) } & & 26.7 & 26.6 & 55.2 &  55.2 &11.9\\
        \hline\hline
    \end{tabular}
    
    \begin{tabular}{cccccccc}
        \hline\hline \text { Capacitance }& & \text { $C_{12}$ } & \text { $C_{13}$ } & \text { $C_{14}$ } & \text { $C_{23}$} & \text { $C_{24}$} & \text { $C_{34}$}  \\
        \hline \text { Value (fF) } & & 0.002 & 12.6 & 0.06 & 0.06 & 12.6 & 30.3 \\
        \hline\hline
    \end{tabular}
    \label{tab:circuit_params}
\end{table}

The frequencies, anharmonicities, and coherence times of the four modes, Q1, Q2, P and M, evaluated experimentally with the coupler biased at the zero-flux point, are provided in Table~S\ref{tab:qubit_params}.

\begin{table}[h]
\caption{Parameters of the four modes, Q1, Q2, P and M, including dressed-state frequencies, anharmonicities and coherence times, i.e., energy-relaxation time $T_1$, Ramsay dephasing time $T_2^\mathrm{R}$, and echo dephasing time $T_2^\mathrm{E}$, when the CSDTC is biased at the zero-flux point.}
    \centering
    \begin{tabular}{cccccc}
        \hline\hline \text { Qubit }& & \text { Q1 } & \text { Q2 } & \text { P } & \text { M}  \\
        \text {Drive frequency (GHz) } & & 3.945 & 4.443  & 6.358 & 5.987 \\
        \text {Anharmonicity (MHz) } & & $-175$ & $-208$ &  & \\
        \text {$T_1$ ($\mathrm{\mu}$s) } & & 232.5 & 161.9 & 37.9 & 59.5 \\
        \text {$T_2^\mathrm{R}$ ($\mathrm{\mu}$s) } & & 299.2 & 171.2 & 66.0 & 85.8 \\
        \text {$T_2^\mathrm{E}$ ($\mathrm{\mu}$s) } & & 316.0 & 202.3 & 65.5 & 90.8 \\
        \hline \hline
    \end{tabular}
    \label{tab:qubit_params}
\end{table}

\section{Condition for suppressed \textit{ZZ} interaction at the zero-flux point }\label{SUP0}

Here, we \Imp{show} the validity of the condition for suppressed \textit{ZZ} interaction  Eq.~(1) derived from the composite impedance of \Imp{$L_{\mathrm{J}5}$} and $C_{34}$ in the main text by employing detailed theoretical models.

\subsection{Derivation with a classical Lagrangian}
\label{secLagrangian}

We start with the following Lagrangian for the CSDTC at \Imp{the} zero-flux point (${\varphi_{\mathrm{ex}}/2\pi=0}$) while ignoring parasitic capacitances for simplicity (${C_{14}=C_{23}=C_{12}=0}$):

\begin{align}
\mathcal{L}=
\sum_{i=1}^4 \frac{C_{ii}}{2} \dot{\Phi}_i^2 
+ \frac{C_{13}}{2} \left( \dot{\Phi}_1 -  \dot{\Phi}_3 \right)^{\! 2}
+ \frac{C_{24}}{2} \left( \dot{\Phi}_2 -  \dot{\Phi}_4 \right)^{\! 2}
+ \frac{C_{34}}{2} \left( \dot{\Phi}_3 -  \dot{\Phi}_4 \right)^{\! 2}
+ \sum_{i=1}^4 E_{\mathrm{J}i} \cos \!\left(\frac{\Phi_i}{\Phi_0} \right)
+ E_{\mathrm{J}5} \cos\!\left(\frac{\Phi_3 - \Phi_4}{\Phi_0}\right),
\end{align}
where $\Phi_i$ and ${E_{\mathrm{J}i} = \Phi_0 I_{\mathrm{c}i}}/2\pi$ are the node flux \Imp{variable} and Josephson energy for JJ$i$, respectively.
By expanding the cosine functions to the fourth order, we obtain
\begin{align}
\mathcal{L}&\simeq \mathcal{L}_{13}+\mathcal{L}_{24}+\mathcal{L}_{34} - V_4,
\\
\mathcal{L}_{13} &= \frac{C_{11}+C_{13}}{2} \dot{\Phi}_1^2 + \frac{C_{11}+C_{13}+C_{34}}{2} \dot{\Phi}_3^2
-C_{13} \dot{\Phi}_1 \dot{\Phi}_3 - \frac{E_{\mathrm{J}1}}{2} \frac{\Phi_1^2}{\Phi_0^2} - \frac{E_{\mathrm{J}3}+E_{\mathrm{J}5}}{2} \frac{\Phi_3^2}{\Phi_0^2},
\\
\mathcal{L}_{24} &= \frac{C_{22}+C_{24}}{2} \dot{\Phi}_2^2 + \frac{C_{22}+C_{24}+C_{34}}{2} \dot{\Phi}_4^2
-C_{24} \dot{\Phi}_2 \dot{\Phi}_4 - \frac{E_{\mathrm{J}2}}{2} \frac{\Phi_2^2}{\Phi_0^2} - \frac{E_{\mathrm{J}4}+E_{\mathrm{J}5}}{2} \frac{\Phi_4^2}{\Phi_0^2},
\\
\mathcal{L}_{34} &= -C_{34} \dot{\Phi}_3 \dot{\Phi}_4 + E_{\mathrm{J}5} \frac{\Phi_3 \Phi_4}{\Phi_0^2},
\\
V_4 &= -\sum_{i=1}^4 \frac{E_{\mathrm{J}i}}{24} \frac{\Phi_i^4}{\Phi_0^4} - \frac{E_{\mathrm{J}5}}{24} \frac{(\Phi_3 - \Phi_4)^4}{\Phi_0^4}.
\end{align}

We transform the flux variables in $\mathcal{L}_{13}$, $\Phi_1$ and $\Phi_3$, as follows: 
\begin{align}
\mathcal{L}_{13} &= \frac{C_1}{2} \dot{\Phi}_1^{\prime 2} + 
\frac{C_3}{2} \dot{\Phi}_3^{\prime 2}
- \frac{E_\mathrm{J}}{2} \frac{\Phi_1^{\prime 2}+\Phi_3^{\prime 2}}{\Phi_0^2},
\label{L13}
\\
\begin{pmatrix}
\Phi'_1 \\ \Phi'_3
\end{pmatrix}
\!
&=
U^{(13) \mathrm{T}} R^{(13)} \!
\begin{pmatrix}
\Phi_1 \\ \Phi_3
\end{pmatrix} 
\!, \,\\
R^{(13)}&=
\!
\begin{pmatrix}
r_{\mathrm{J}1} & 0 \\ 0 & r_{\mathrm{J}3}
\end{pmatrix}
\!, \,\\
r_{\mathrm{J}1}&=\sqrt{\frac{E_{\mathrm{J}1}}{E_\mathrm{J}}},  
\,\\
r_{\mathrm{J}3}&=\sqrt{\frac{E_{\mathrm{J}3}+E_{\mathrm{J}5}}{E_\mathrm{J}}},
\\
U^{(13) \mathrm{T}} & R^{(13)-1}
\!
\begin{pmatrix}
C_{11}+C_{13} & -C_{13} \\ -C_{13} & C_{33}+C_{13}+C_{34}
\end{pmatrix}
\!
R^{(13)-1} U^{(13)}=
\!
\begin{pmatrix}
C_1 & 0 \\ 0 & C_3
\end{pmatrix}
\!,
\end{align}
where $E_\mathrm{J}$ is an arbitrary energy, e.g.\ $E_{\mathrm{J}1}$, for normalization, 
$C_1$ and $C_3$ are the eigenvalues of the normalized capacitance matrix for $\mathcal{L}_{13}$, 
$U^{(13)}$ is the unitary matrix composed of the corresponding eigenvectors, and 
$U^{(13) \mathrm{T}}$ is the transpose of $U^{(13)}$.
We also transform the flux variables in $\mathcal{L}_{24}$, $\Phi_2$ and $\Phi_4$, similarly, 
and obtain the following:
\begin{align}
&
\mathcal{L}_{24} = \frac{C_2}{2} \dot{\Phi}_2^{\prime 2} + 
\frac{C_4}{2} \dot{\Phi}_4^{\prime 2}
- \frac{E_\mathrm{J}}{2} \frac{\Phi_2^{\prime 2}+\Phi_4^{\prime 2}}{\Phi_0^2}.
\label{L24}
\end{align}
Note that $\Phi'_1$ and $\Phi'_2$ correspond to the two qubits.

By the above transformations, $\mathcal{L}_{34}$ approximately gives the following interaction Lagrangian between the two data qubits:
\begin{align}
\mathcal{L}_{12}
=
\frac{U_{21}^{(13)} U_{21}^{(24)}}{r_{\mathrm{J}3} r_{\mathrm{J}4}} \!
\left(
 -C_{34} \dot{\Phi}'_1 \dot{\Phi}'_2 + E_{\mathrm{J}5} \frac{\Phi'_1 \Phi'_2}{\Phi_0^2}
\right).
\end{align}
From Eqs.~(\ref{L13}) and (\ref{L24}) \Imp{and assuming that the interaction and detuning between the two qubits are sufficiently small, 
$\Phi'_i$ is approximately expressed as ${\Phi'_i (t)} = {\Phi_i^{(0)} \! (t)} {\cos (\omega_i t + \theta_i)}$ ($i=1, 2$), 
where ${\Phi_i^{(0)} \! (t)}$ is a slowly varying amplitude, 
${\omega_i = 1/\sqrt{L_\mathrm{J} C_i}}$ (${L_\mathrm{J} = \Phi_0^2/E_\mathrm{J}}$), and $\theta_i$ is a phase.
Dropping ${\dot{\Phi}_i^{(0)} \! (t)}$ and in the rotating-wave approximation, the interaction Lagrangian $\mathcal{L}_{12}$ becomes}
\begin{align}
\mathcal{L}_{12}
=
\frac{U_{21}^{(13)} U_{21}^{(24)}}{r_{\mathrm{J}3} r_{\mathrm{J}4}} \!
\left(
 -C_{34} \omega_1 \omega_2 
+ \frac{E_{\mathrm{J}5}}{\Phi_0^2} 
\right) \!
{\Phi_1^{(0)} \! (t)} {\Phi_2^{(0)} \! (t)} 
\frac{\cos [(\omega_1-\omega_2) t + \theta_1 - \theta_2]}{2}.
\end{align}

Thus, \Imp{neglecting $V_4$}, 
the condition for turning off the qubit--qubit coupling is given by
\begin{align}
 -C_{34} \omega_1 \omega_2 
+ \frac{E_{\mathrm{J}5}}{\Phi_0^2} 
=0 
\iff
\frac{1}{\sqrt{L_{\mathrm{J}5} C_{34}}}
=
\sqrt{\omega_1 \omega_2},
\end{align}
which is exactly the same as that derived from the composite impedance of \Imp{$L_{\mathrm{J}5}$} and $C_{34}$ in the main text.

\subsection{Derivation with a quantum-mechanical Hamiltonian}

The Hamiltonian corresponding to the Lagrangian in Sec.~\ref{secLagrangian} approximately becomes as follows~\cite{PhysRevApplied.18.034038}:
\begin{align}
\mathcal{H}
=
4\hbar \mathbf{n}^\mathrm{T} W \mathbf{n} + 
\sum_{i=1,2} \left[ 
\frac{E_\mathrm{J}}{2} \varphi_i^2
- \frac{E''_{\mathrm{J}i}}{24} \varphi_i^4 \right]
- E'_{\mathrm{J}5} \varphi_1 \varphi_2
- \frac{E''_{\mathrm{J}5}}{4} \varphi_1^2 \varphi_2^2,
\label{eqHamiltonian}
\end{align}
where ${\varphi_i = \Phi'_i/\Phi_0}$ is the node phase, and 
the Cooper-pair numbers ${\mathbf{n}=(n_1~ n_2)^ \mathrm{T}}$ are defined as 
\begin{align}
\mathbf{n}&=
\frac{1}{2e} M_\mathrm{c} 
\!
\begin{pmatrix}
\dot{\Phi}'_1 \\ \dot{\Phi}'_2
\end{pmatrix}
\!, \\
M_\mathrm{c}&=
\!
\begin{pmatrix}
C_1 & -C'_{34} \\ -C'_{34} & C_2
\end{pmatrix}.
\end{align}
The other parameters are defined as follows:
\begin{align}
\hbar W &= \frac{e^2}{2} M_\mathrm{c}^{-1},
\\
C'_{34} &= k_{\mathrm{Ur}} C_{34}, \label{Cd34}
\\
k_{\mathrm{Ur}} &= \frac{U_{21}^{(13)} U_{21}^{(24)}}{r_{\mathrm{J}3} r_{\mathrm{J}4}},
\\
E''_{\mathrm{J}1} &= E_{\mathrm{J}1} \! \left( \frac{U_{11}^{(13)}}{r_{\mathrm{J}1}} \right)^{\! 4},
\\
E''_{\mathrm{J}2} &= E_{\mathrm{J}2} \! \left( \frac{U_{11}^{(24)}}{r_{\mathrm{J}2}} \right)^{\! 4},
\\
E'_{\mathrm{J}5} &= k_{\mathrm{Ur}} E_{\mathrm{J}5},
\label{EdJ5}
\\
E''_{\mathrm{J}5} &= k_{\mathrm{Ur}}^2 E_{\mathrm{J}5}.
\end{align}

The Hamiltonian in Eq.~(\ref{eqHamiltonian}) is quantized as
\begin{align}
\mathcal{H}
&=
\hbar \omega_1 a_1^{\dagger} a_1 + \hbar \omega_2 a_2^{\dagger} a_2 
+ g_{12} \! \left( a_1^{\dagger} a_2 +a_2^{\dagger} a_1 \right) + V'_4,
\\
\varphi_i &= \sqrt{\frac{8W_{ii}}{\omega_i}}\, \frac{a_i + a_i^{\dagger}}{\sqrt{2}}, 
\\
n_i &= \sqrt{\frac{\omega_i}{8W_{ii}}}\, \frac{a_i - a_i^{\dagger}}{\sqrt{2}i}, 
\end{align}
where ${\omega_i = \sqrt{8W_{ii} w_\mathrm{J}}}$ (${\omega_J=E_\mathrm{J}/\hbar}$), $a_i^{\dagger}$ and $a_i$ are the creation and annihilation operators for Cooper pairs on node~$i$, 
the rotating-wave approximation has been made, $V'_4$ denotes the fourth-order terms, 
and the effective coupling rate $g_{12}$ between the qubits is defined as follows:
\begin{align}
g_{12} = \frac{1}{2} \sqrt{\frac{\omega_1 \omega_2}{W_{11} W_{22}}}
\! \left(
W_{12} - 8 \frac{E'_{\mathrm{J}5}}{\hbar} \frac{W_{11} W_{22}}{\omega_1 \omega_2}
\right).
\end{align}

Using $W_{12} \simeq [e^2/(2\hbar)] [C'_{34}/(C_{11} C_{22})] =8W_{11} W_{22} C'_{34} \Phi_0^2/\hbar$, 
Eqs.~(\ref{Cd34}) and (\ref{EdJ5}), and $E_{\mathrm{J}5}=\Phi_0^2/L_{\mathrm{J}5}$, 
\begin{align}
g_{12} = 0
\iff
C_{34} = \frac{1}{L_{\mathrm{J}5} \omega_1 \omega_2}, 
\end{align}
which is again exactly the same as that derived from the composite impedance of \Imp{$L_{\mathrm{J}5}$} and $C_{34}$ in the main text.

The \Imp{\textit{ZZ}} \Imp{interaction} $\zeta$ can also be formulated as follows.
First, we transform the two modes to the eigenmodes including the interaction as follows:
\begin{align}
&
\begin{pmatrix}
b_1 \\ b_2 
\end{pmatrix}
\!
=
U^{(12)\mathrm{T}}
\!
\begin{pmatrix}
a_1 \\ a_2 
\end{pmatrix}
\!,
\, \\
&
U^{(12)\mathrm{T}}
\!
\begin{pmatrix}
\omega_1 & g_{12} \\ g_{12} & \omega_2 
\end{pmatrix}
\!
U^{(12)}
=
\!
\begin{pmatrix}
\omega'_1 & 0 \\ 0 & \omega'_2 
\end{pmatrix}
\!,
\end{align}
where $b_1$ and $b_2$ are the annihilation operators for the eigenmodes,  
$\omega'_1$ and $\omega'_2$ are the corresponding eigenfrequencies, 
and $U^{(12)}$ is the unitary matrix composed of the eigenvectors.
By substituting the relation into $V'_4$, 
the \Imp{\textit{ZZ}} \Imp{interaction} $\zeta$ (cross-Kerr coefficient) is 
identified as the coefficient of ${b_1^{\dagger} b_1 b_2^{\dagger} b_2}$ as follows \Imp{(${\omega''_{\mathrm{J}i}=E''_{\mathrm{J}i}/\hbar}$)}:
\begin{align}
\zeta =
-\frac{\omega''_{\mathrm{J}1}}{4} \! \left( \frac{8W_{11}}{\omega_1} \right)^{\!2} U_{11}^{(12)2} U_{12}^{(12)2}
-\frac{\omega''_{\mathrm{J}2}}{4} \! \left( \frac{8W_{22}}{\omega_2} \right)^{\!2} U_{21}^{(12)2} U_{22}^{(12)2}
-\frac{\omega''_{\mathrm{J}5}}{4} \frac{8W_{11}}{\omega_1} \frac{8W_{22}}{\omega_2} U_{11}^{(12)2} U_{22}^{(12)2}. 
\label{eq-ZZ}
\end{align}

\begin{figure}[h]
	\includegraphics[width=8cm]{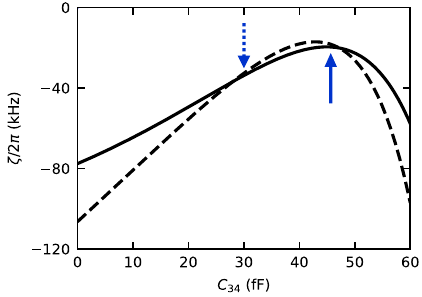}
	\caption{\textit{ZZ} \Imp{interaction} $\zeta$ as a function of the shunt capacitance $C_{34}$. Solid and dashed curves are $\zeta$ obtained 
	by Eq.~(\ref{eq-ZZ}) and numerical diagonalization without approximations, respectively.
	Solid and dotted arrows indicate our formula ${C_{34} = 1/(L_{J5} \omega_1 \omega_2)}$ and $C_{34}$ in our experiment, respectively.}
	\label{FigS2:wide}
\end{figure}

To verify the above approximations, 
we numerically compare the \Imp{\textit{ZZ}} \Imp{interaction} $\zeta$ using Eq.~(\ref{eq-ZZ}) with 
that calculated by numerical diagonalization of the Hamiltonian without approximations~\cite{PhysRevApplied.18.034038}.
The result is shown in Fig.~S\ref{FigS2:wide}, where the circuit parameters are set to the values reproducing the experimental data 
except for $C_{14}=C_{23}=C_{12}=0$ and varying $C_{34}$. 
Thus, our formula well reproduces the actual $\zeta$, in particular, around the minimum-\textit{ZZ}-interaction point.

\section{Error budget of the CZ gate}\label{SUP1}

Here, we employ \Imp{leakage randomized benchmarking (LRB)} and purity randomized benchmarking (PRB) methods to systematically decompose the CZ-gate error into its constituent components: incoherent, coherent, and leakage errors.
\begin{figure}
\includegraphics[scale=1.0]{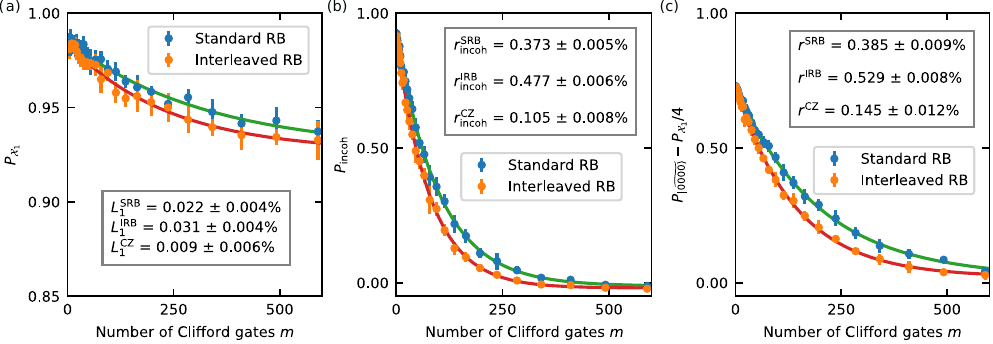}% Here is how to import EPS art
\caption{\label{FigS3:wide} Interleaved randomized benchmarking for (a)~leakage error~$L_1^{\mathrm{CZ}}$, (b)~incoherent error~$r^\mathrm{CZ}_\mathrm{incoh}$ and (c)~CZ-gate error $r^\mathrm{CZ}$.}
\end{figure}

\subsection{Leakage error}\label{SUP1-1}
By using the population in the computational subspace, $\mathcal{X}_1 = \{|\widetilde{0000}\rangle, |\widetilde{0100}\rangle, |\widetilde{1000}\rangle, |\widetilde{1100}\rangle\}$ measured during the interleaved randomized benchmarking, a leakage error $L_1^{\mathrm{CZ}}$ can be calculated~[Fig.~S\ref{FigS3:wide}(a)]~\cite{PhysRevLett.123.120502}. 
The resulting curves are fitted to
\begin{equation}
P_{\mathcal{X}_1}(m)=A + B \lambda_L^{m},
\end{equation}
where $m$ is the Clifford-gate number in one sequence.
For standard randomized benchmarking~(SRB) and interleaved randomized benchmarking~(IRB), the leakage error $L_1^\mathrm{SRB(IRB)}$ for an average Clifford gate is estimated as
\begin{equation}
L_1^\mathrm{SRB(IRB)}= (1 - A)(1 - \lambda_L^\mathrm{SRB(IRB)}).
\end{equation}
Then the leakage error of the CZ gate can be calculated as
\begin{equation}
    L_1^{\mathrm{CZ}}=1-\frac{1-L_1^\mathrm{IRB}}{1-L_1^\mathrm{SRB}}.
\end{equation}

\subsection{Incoherent error}\label{SUP1-2}
Purity randomized benchmarking~\cite{wallman2015estimating} is employed to estimate the incoherent error $r_\mathrm{incoh}^{\mathrm{CZ}}$, as illustrated in Fig.~S\ref{FigS3:wide}(b). Initially, the state population $\rho_{\chi_1}$ is reconstructed through tomography, and then the normalized purity~\cite{PhysRevLett.117.260501} is modeled as
\begin{equation}\label{Purity RB}
\begin{aligned}
P_{\mathrm{incoh}} (m) & =\frac{d}{d-1}\left[\mathrm{tr}\left[\rho_{\chi_1}^2 (m)\right] - \frac{1}{d}\right]\\
& = A \lambda_\mathrm{incoh}^{2m} + B,
\end{aligned}
\end{equation}
as a function of the sequence length $m$. The dimension $d = 4$ represents the  two-qubit computational subspace. Therefore, the incoherent error $r_\mathrm{incoh}^{\mathrm{CZ}}$ is calculated as
\begin{equation}\label{incoherent error}
r_\mathrm{incoh}^{\mathrm{CZ}} \equiv 1 - \frac{1-r_\mathrm{incoh}^{\mathrm{IRB}}}{1-r_\mathrm{incoh}^{\mathrm{SRB}}}\approx \frac{d-1}{d}\left(1 - \frac{ \lambda_\mathrm{incoh}^\mathrm{IRB}}{ \lambda_\mathrm{incoh}^\mathrm{SRB}}\right), 
\end{equation}
where
\begin{equation}\label{incoherent r}
r_\mathrm{incoh}^{\mathrm{SRB(IRB)}} \equiv (1-\lambda_\mathrm{incoh}^\mathrm{SRB(IRB)})(1-1/d).
\end{equation}
We note that a maximum-likelihood estimation~\cite{PhysRevA.84.042108} can be employed to reconstruct the quantum state, under the intrinsic assumption of no leakage during quantum state tomography. For simplicity, however, we directly construct the state density matrix, which results in the baseline of the normalized purity values falling below zero due to the leakage, as seen in Fig.~S\ref{FigS3:wide}(b). Additionally, we neglect the contribution of leakage errors in fitting the PRB curves, as they are an order of magnitude smaller than incoherent errors in our study. While this approximation may slightly affect the precise analysis of the CZ-gate error budget, it does not alter the CZ-gate fidelity we estimated.

\subsection{Coherent error}\label{SUP1-3}
We use the method detailed in Ref.~\citenum{PhysRevX.14.041050} to extract the coherent error using the IRB method. By measuring the population of $|\widetilde{0000}\rangle$, $P_{|\widetilde{0000}\rangle}$, a gate error $r^\mathrm{CZ}$ can be estimated in combination with the leakage error benchmarking results~[Fig.~S\ref{FigS3:wide}(c)]. The curves are fitted to
\begin{equation}\label{P00-fit}
    P_{|\widetilde{0000}\rangle}(m) - P_{\mathcal{X}_1}(m)/d = C\lambda_r^{m} +D.
\end{equation}
Then, $r^\mathrm{CZ}$ can be calculated as
\begin{equation}
    r^\mathrm{CZ} =1-\frac{1-r^\mathrm{IRB}}{1-r^\mathrm{SRB}}\approx \frac{(d-1)}{d}\left(1-\frac{\lambda_r^\mathrm{IRB}}{ \lambda_r^\mathrm{SRB}}\right),
\end{equation}
with $r^\mathrm{SRB(IRB)} \equiv (1-\lambda_r^\mathrm{SRB(IRB)})(1-1/d)$. Note that $r^\mathrm{CZ} = r_\mathrm{incoh}^{\mathrm{CZ}} + r_\mathrm{coh}^{\mathrm{CZ}} + \frac{3}{4}L_1^{\mathrm{CZ}}$~\cite{PhysRevX.14.041050}, where $r_\mathrm{coh}^{\mathrm{CZ}}$ represents the coherent error. Therefore, the coherent error can be estimated as $r_\mathrm{coh}^{\mathrm{CZ}} = r^\mathrm{CZ} -( r_\mathrm{incoh}^{\mathrm{CZ}} +  \frac{3}{4}L_1^{\mathrm{CZ}})$.

The total gate error can be well characterized as $1-F^{\mathrm{CZ}} = r^{\mathrm{CZ}} + \frac14 L_1^{\mathrm{CZ}} = r_\mathrm{incoh}^{\mathrm{CZ}} + r_\mathrm{coh}^{\mathrm{CZ}} + L_1^{\mathrm{CZ}}$~\cite{PhysRevA.97.032306, PhysRevX.14.041050}.

\section{Incoherent error for BNZ and SP pulses}\label{SUP2}

\begin{figure}
\includegraphics[scale=1.0]{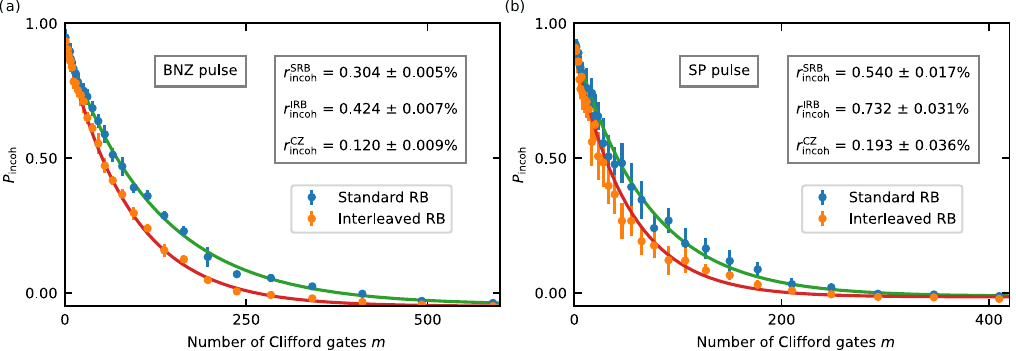}% Here is how to import EPS art
\caption{\label{FigS4:wide} Purity randomized benchmarking of the CZ gates with  (a)~biased net-zero~(BNZ) pulse and (b)~single-pole~(SP) pulse, which correspond to the pulses shown in Fig.~4(a) in the main text.}
\end{figure}

When the CSDTC is biased at $\varphi_\mathrm{ex}/2\pi = 0.2$, the BNZ pulse achieves higher CZ-gate fidelity than the SP pulse, as characterized in Fig.~4(b) of the main text. Here, we use PRB to identify the dominant error in each case (Fig.~S\ref{FigS4:wide}). For the BNZ pulse, incoherent error accounts for 86.4\% of the total error, while for the SP pulse, it contributes 57.5\% of the total error. In each case, the leakage error contribution is below 10\%,  indicating that coherent error constitutes a significant portion of the total error in the SP-pulse CZ gate. Furthermore, we estimate that the incoherent error induced by 1/\textit{f} flux noise remains below 
$4\times 10^{-6}$ in both cases, consistent with the previous results~\cite{PhysRevX.14.041050}.

\renewcommand\refname{Reference}

\bibliography{s_ref.bib}% Produces the bibliography via BibTeX.